# Real-space imaging of a topological protected edge state with ultracold atoms in an amplitude-chirped optical lattice

Martin Leder[1], Christopher Grossert[1], Lukas Sitta[1], Maximilian Genske[2], Achim Rosch[2], and Martin Weitz[1]

[1]*Institut für Angewandte Physik, Universität Bonn, Wegelerstr. 8, 53115 Bonn, Germany*

[2]*Institut für Theoretische Physik, Universität zu Köln, Zülpicher Str. 77, 50937 Cologne, Germany*

**To describe a mobile defect in polyacetylene chains, Su, Schrieffer and Heeger formulated a model assuming two degenerate energy configurations that are characterized by two different topological phases. An immediate consequence was the emergence of a soliton-type edge state located at the boundary between two regions of different configurations. Besides giving first insights in the electrical properties of polyacetylene materials, interest in this effect also stems from its close connection to states with fractional charge of relativistic field theory. Using a one-dimensional optical lattice for cold rubidium atoms with a spatially chirped amplitude, we experimentally realize an interface between two spatial regions of different topological order in an atomic physics system. We directly observe atoms confined in the edge state at the intersection by optical real-space imaging and characterize the state as well as the size of the associated energy gap. Our findings hold prospects for the spectroscopy of surface states in topological matter and for the quantum simulation of interacting Dirac systems.**

**Introduction**

Topological states of matter, as quantum Hall systems or topological insulators, cannot be distinguished from ordinary matter by local measurements in the bulk of the material[1-4].

Instead, global measurements are required, revealing topological invariants as the Chern number. At the heart of topological materials are topologically protected edge states that occur at the intersection between regions of different topological order[3-7]. Ultracold atomic gases in optical lattices are promising new platforms for topological states of matter[8-13], though the observation of edge states has so far been restricted in these systems to the state space imposed by the internal atomic structure[14,15].

Here we report on the observation of an edge state between two topological distinct phases of an atomic physics system in real space using optical microscopy. An interface between two spatial regions of different topological order is realized in a one-dimensional optical lattice of spatially chirped amplitude. To reach this, a magnetic field gradient causes a spatial variation of the Raman detuning in an atomic rubidium three-level system and a corresponding spatial variation of the coupling between momentum eigenstates. This novel experimental technique realizes a cold atom system described by a Dirac equation with an inhomogeneous mass term closely related to the Su, Schrieffer and Heeger (SSH) model[16,17]. The observed edge state is characterized by measuring the overlap to various initial states, revealing that this topological state has singlet nature in contrast to the other system eigenstates, which occur pairwise. We also determine the size of the energy gap to the adjacent eigenstate doublet.

**Results**

*Background*

Fig.1a shows a schematic of the spatial variation of the relevant band structure of rubidium atoms in the one-dimensional lattice. The lattice has a spatial periodicity of $\lambda/4$, where $\lambda$ denotes the wavelength of the driving laser beams, with the corresponding potential being due to the dispersion of Raman transitions[18,19]. A magnetic field gradient imprints a spatially slowly varying lattice amplitude (in comparison with the scale of the lattice periodicity) via a modification of the Raman detuning. The achieved coupling K(z) between momentum eigenstates $\exp(\pm 2ikz)$ determining the lattice amplitude, with $k = 2\pi/\lambda$, varies spatially along the z-axis. For not too large values of z the coupling follows $K(z) = a \cdot z$, where $a$ is a constant, and thus changes sign from negative to positive values of z. Here $K(z)$ varies only slowly in space and it is therefore useful to discuss the bandstructure for fixed $K(z) \approx K(z_0)$. At position $z_0 = 0$, where the coupling $K$ between momentum eigenstates vanishes, a crossing of Bloch bands occurs and the ordering of bands, indicated as |-> and |+> respectively in Fig.1a, is inverted. By continuous deformation the bands cannot be transformed into each other without closing the gap between bands. For such a situation a topologically protected edge state, localized around z=0 where the bands intersect, is expected[20,21]. Formally, this can be seen with a simple model describing the system near the crossing by a one-dimensional Dirac-Hamiltonian[22,23] with a spatially dependent effective mass $m_{eff}(z) = a \cdot z / c_{eff}^2$:

$$\hat{H}_D = m_{eff}(z) c_{eff}^2 \sigma_x + \hat{q} c_{eff} \sigma_z, \qquad (1)$$

where $\hat{q} = -i\hbar \partial_z$ is the momentum operator, $\sigma_x$ and $\sigma_z$ are Pauli matrices, and $c_{eff} = 2\hbar k / m \cong 1.1 \text{cm/s}$ is an effective speed of light, with $m$ as the rubidium atomic mass. The two-component Hamiltonian acts on spinors $\psi(z) = (\Psi_1(z), \Psi_2(z))$, with $\Psi_1$ and $\Psi_2$ corresponding to wavefunctions of atoms with momenta close to $\pm 2\hbar k$, respectively. The total wavefunction is given by $\phi(z) = \Psi_1(z) e^{2ikz} + \Psi_2(z) e^{-2ikz}$. The eigenenergies of the system are readily found to be $E_0 = 0$ and $E_n = \pm\sqrt{n} \cdot \hbar \omega_0$ for n>0, with $\omega_0 = \sqrt{2 c_{eff} a / \hbar}$. For quantum numbers n>0 the solution comes in pairs of opposite energies, while for n=0 there is only a singlet eigenstate, the topological edge state at the energetic position of the band crossing, which originates from the vanishing energy gap at the interface. The wavefunction of this state is

$$\psi_0(z) = \frac{1}{\sqrt{2}} \left( \frac{a}{\pi c_{\text{eff}} \hbar} \right)^{1/4} \exp\left(-az^2/(2c_{\text{eff}}\hbar)\right) \cdot \begin{pmatrix} 1 \\ -i \end{pmatrix}, \qquad (2)$$

which is the product of a Gaussian envelope and a coherent superposition of the two momentum eigenstates with a relative phase of $\pi/2$. As suggested by the discrete nature of the eigenenergies, all system eigenstates are bound states. Fig.1b shows the spatial variation of the probability density for the topological edge state (middle) and the doublet states with $n=1$, 2 and 3 (the corresponding plots on the top and bottom respectively).

*Experimental realization*

Our experiment starts by initially cooling a dilute cloud of rubidium atoms ($^{87}$Rb) in the $m_F=-1$ spin projection of the F=1 hyperfine component to Bose-Einstein condensation in a combined optical dipole and magnetic trap. A magnetic field gradient of magnitude $m \cdot g /(\mu_B/2) \cong 30.5\,\text{G}\,\text{cm}^{-1}$ compensates for the Earth's gravitational acceleration. The cold atomic cloud is subsequently adiabatically expanded along the z-axis to match the size of the topological edge state, of expected width $\Delta z = \sqrt{c_{\text{eff}}\hbar/2a} \cong 11.5\,\mu\text{m}$, in order to obtain sufficient spatial overlap. The typical experimental width of the momentum distribution at this point is $\Delta p_z \cong 0.01\hbar k$, corresponding to an effective kinetic temperature of 35pK.

The lattice potential is realized using a rubidium atomic three-level configuration with two ground states of different spin projections and one spontaneously decaying excited state[19], see Fig.2a. To achieve a zero crossing of the coupling between bands, two four-photon potentials $V_1(z)$, $V_2(z)$ with opposite spatial variation of the coupling are superimposed, as realized by choosing values of the Raman detuning $\delta_{1,2} = \pm \delta_0 - \mu_B \Delta B(z)/(2\hbar)$ of opposite signs, where $\delta_0$ is a constant and $\Delta B(z) = B(z) - B(0) \cong z \cdot dB/dz$ with the spatially dependent magnetic $B(z)$ field tuning the Zeeman splitting. We arrive at a lattice potential $V(z) = \frac{1}{2}(V_{1,0}(z) + V_{2,0}(z))\cos(4kz)$ with the envelopes $V_{i,0}(z) = \hbar\Omega_{\text{eff}}^+\Omega_{\text{eff}}^- / \delta_i(z)$, where $\Omega_{\text{eff}}^+$ and $\Omega_{\text{eff}}^-$ denote two-photon Rabi frequencies[18]. Expanding the amplitude of the lattice potential to first order around z=0 leads to $V_{i,0}(z) = \pm V_0 + 2a \cdot z$, with $V_0 = \hbar\Omega_{\text{eff}}^+\Omega_{\text{eff}}^- / \delta_0 \cong 0.36 \cdot E_r$, where $E_r = \hbar^2 k^2 /(2m)$ denotes the recoil energy, and $a = V_0 \mu_B (dB/dz)/(4\hbar\delta_0) \cong 19.0 \cdot E_r$ cm$^{-1}$. The total lattice potential can be written in the simple form $V(z) = 2a \cdot z \cos(4kz)$, which experiences a zero crossing at z=0. Note that for z>0 (z<0) the maxima (minima) of the potential are located at integer multiples of $\lambda/4$, see Fig.2b. This phase change is reflected in the inversion of band ordering. The dynamics of atoms in such a structure near the band crossing is described by the Dirac Hamiltonian of eq.1 with good accuracy (see methods), as the width of the topological bound state $\Delta z$ is 2 orders of magnitude larger than the lattice spacing.

*Characterization of the spatial variation of the band structure*

In preparatory experiments we have characterized the band structure of the one-dimensional lattice focusing on the band-inversion upon sign change of z. For this, the adiabatically expanded atomic cloud centered at different lateral positions $z_0$ along the lattice beam axis (described by a wavefunction $\phi_a(z - z_0)$) is transferred to the state $\phi_i(z) = \frac{1}{\sqrt{2}}\phi_a(z - z_0)\left(e^{i2kz} + e^{-i2kz}\right)$ via two simultaneously performed Bragg pulses[24], described by the spinor $\psi_i(z) = \phi_a(z - z_0)|+\rangle$ with $|\pm\rangle = \begin{pmatrix} 1 \\ \pm 1 \end{pmatrix}/\sqrt{2}$, see Fig.1. To verify whether the cloud at the corresponding position overlaps with the upper or the lower band in the lattice, respectively, the band populations following activation of the lattice are determined. As shown in Fig.3, for the chosen initial state we find that for $z_0 < 0$ the loading is enhanced into the lower band, while for $z_0 > 0$ most atoms are transferred into the upper band, and near $z_0=0$ the curves cross. This experimentally verifies the expected spatial variation of the band structure, exhibiting a sign change of the coupling between momentum eigenstates at $z_0=0$.

*Imaging cold atoms at the topological interface*

We have next loaded atoms into the topological edge state. For this, using the Bragg-pulses atoms are transferred to an initial state $\phi_i(z) = \frac{1}{\sqrt{2}} \phi_a(z)\left(e^{i2kz} + e^{-i\varphi}e^{-i2kz}\right)$ with $\varphi=\pi/2$. The atomic wavepacket is centered at $z=0$, after which the lattice beams are activated. As the Bragg pulse increases the kinetic energy by more than two orders of magnitude, while the interaction energy remains largely unaffected, it turns out (see methods) that interactions can be ignored for the subsequent time evolution.

Fig.4a (top) shows a series of atomic absorption images recorded after a variable holding time in the lattice along with a simulation (bottom). We observe that the atomic cloud remains trapped at the expected position of the atomic edge state. On the other hand, for a relative phase of $\varphi=-\pi/2$ no such trapping in the edge state is observed, see Fig.4b (top). This is in agreement with expectations, as when the initially prepared atomic wavepacket is $\pi$ out of phase there is no overlap with the topological edge state. Instead, the wavepacket is split up into two spatially diverging paths. For larger times, the onset of an oscillation is visible, which during the experimentally accessible interaction times near 2ms is only partially resolved. A fit yields a period $T = 3.16(32)\,\text{ms}$, see Methods. Such an oscillatory motion is also seen in the simulations (Fig.4b, bottom). From theory we expect that for $\varphi=\pi/2$ the loading efficiency into the topological state is 95%, while for $\varphi=-\pi/2$ the wavepacket is mainly described by a coherent superposition of the two eigenstates of the first doublet with $n=1$, which beat with an oscillation period of $T=\pi/\omega_0$. Our experimental oscillation data allows us to determine the size of the splitting $\omega_0/2\pi$ to $158(16)\,\text{Hz}$, which is in good agreement with the expected value of $163\,\text{Hz}$, and gives a direct measurement for the size of the gap between the topologically protected edge state and the two energetically closest other system eigenstates. The observed lifetime near 2ms is attributed to photon scattering from the Raman beams, an effect also assigned to be dominantly responsible for an observed residual expansion of the edge state visible in Fig.4a (top). A further contribution to the observed residual expansion is a remaining mismatch of the initially prepared atomic wavepacket with the topological edge state, causing an admixture of eigenstates with larger values of $n$ and correspondingly increased mode volume. This effect was accounted for in the model simulations shown in Fig.4 (bottom), see methods.

*Detailed characterization: phase dependency and loading efficiency*

Fig.5a shows a series of absorption images recorded after a fixed time $t_{\text{int}}$=1.7ms for different values of the relative phase between momentum components of the initially prepared atomic wavepacket. Near a relative phase of $\varphi=\pi/2$ we again observe a compact atomic cloud, while for $\varphi=-\pi/2$ the cloud is split up into two components. A smooth variation between these extremes is visible for intermediate phase values. The corresponding variation of the total rms width of the atomic cloud along the z-axis versus the phase φ is shown in Fig.5b.

We have next modified the overlap of the initial state and the topological state by preparing clouds with smaller initial size and correspondingly larger momentum spread by using different final values of the dipole trapping potential during adiabatic expansion. As described above, adiabatic expansion of the condensate cloud yields atomic ensembles with down to $0.01\hbar k$ momentum width, corresponding to effective temperatures in the pK regime. The dots in Fig.5c show the relative variation of the total cloud width on the phase versus the momentum width of the atomic cloud. For a larger momentum spread of the initial wavepacket the observed phase dependency of the total cloud size reduces. This is attributed to the higher order system eigenstates with $n>0$, which are populated when loading with a wavepacket of larger momentum width than of the topological protected edge state. The experimental results are in agreement with a simulation (solid line). The inset of Fig.5c shows the expected variation of the loading efficiency into the edge state on the momentum width of the cloud.

**Discussion**

To conclude, an edge state at the spatial interface between two regions of different topological order has been observed by real-space imaging of the cold atomic cloud. Evidence for the successful population of this topological state has been obtained from (i) the phase dependence, (ii) the dependence of the initial atomic momentum width of the loading efficiency and (iii) the vibrational frequency of the lowest excited doublet modes agreeing with expectations.

For the future, it will be important to reveal the role of interactions on the topological edge state using e.g. Feshbach resonances for sensitive control, with prospects including the simulation of interacting topological quantum matter[25]. Other perspectives include the realization of interacting relativistic wave equation predictions[26], as well as novel topological Berry phase effects in phase space[27].

**Acknowledgements**

We acknowledge financial support by the DFG (We 1748-20), the Bonn-Cologne Graduate School and the Deutsche Telekom Stiftung (M.G.).

**Figure Captions:**

**Figure 1 | Band structure and eigenstates. a,** Illustration of spatial variation of the band structure of the lowest two bands denoted as |+> and |->, respectively, along position *z*. The edge state, indicated by the green line at *z*=0, is located between two regions of different band ordering defining the topology, with *z*<0 and *z*>0 respectively. **b**, Probability distribution of eigenstates versus position z in units of $\Delta z = \sqrt{c_{\text{eff}} \hbar/2a}$, for the states with quantum numbers $n \leq 3$. The middle green plot refers to the topological edge state with *n*=0, which is a singlet. All other eigenstates are doublets, see the plots on the top (blue) and bottom (red), with energies $E = \hbar \omega_0 \sqrt{n}$ and $E = -\hbar \omega_0 \sqrt{n}$, respectively. For the explicit analytical form of the wavefunction of the states with *n*>0, see Methods.

**Figure 2 | Experimental schematic. a**, Four-photon scheme for generation of amplitude-chirped lattice potentials $V_1(z)$ (top panel) and $V_2(z)$ (bottom panel). The scheme is based on three-level atoms with ground state sublevels |$m_F$=-1> and |$m_F$=0>, and the excited manifold |e>, driven by an optical field of frequency $\omega$ and two counterpropagating fields of frequencies $\omega \pm \Delta \omega_{1,2}$. A magnetic field gradient causes a spatial variation of the Raman-detuning $\delta_{1,2}(z)$. **b**, The plot for $V_1(z)$ ($V_2(z)$) indicates the spatial variation of the lattice depth, which increases (decreases) along z for positive (negative) values of the Raman detuning. A lattice with band structure as shown in Fig.1a is obtained by superimposing $V_1(z)$ and $V_2(z)$. The bottom panel gives a schematic of the setup. In the real experiment, the magnetic field gradient is directed vertically.

**Figure 3 | Spatial variation of band ordering**. Relative atomic population transferred into the upper (green circles) and lower (blue squares) band upon loading from an initial state $\phi_i(z) = \frac{1}{\sqrt{2}} \phi_a(z - z_0)(e^{i2kz} + e^{-i2kz})$ with an atomic cloud centered at position $z_0$. The band

populations were determined by accelerating the lattice away from the crossing to a relative wavevector of $0.5\hbar k$ so that a mapping onto the free atomic eigenstates occurs, and then applying time-of-flight imaging[24]. Each point corresponds to the average of three corresponding data sets, and the shown error bars are the standard deviation of the mean.

**Figure 4 | Temporal evolution of atomic clouds**. Series of absorption images (top) for a relative phase of the initially prepared atomic wavepacket of **(a)** $\varphi=\pi/2$ and **(b)** $\varphi=-\pi/2$, respectively, for different holding times in the lattice. The temporal step size between images is 0.1ms, and the measured optical column density is encoded in color code. For $\varphi=\pi/2$ we observe a trapping of atoms in the topological edge state, while for $\varphi=-\pi/2$ the cloud splits up. Each image is the average over four realizations. The bottom plots are numerical simulations taking into account the experimental resolution of 4.8μm (see Methods).

**Figure 5 | Phase and momentum width dependence of loading. a**, Series of absorption images for different relative phases of the Raman state preparation beams. **b**, Corresponding variation of the rms widths along the z-axis of the distribution (blue dots) along with a sinusoidal fit (solid line). Each data point is the average over four realizations, and the shown error bars correspond to the standard deviation of the mean. The green squares are the result of a second data set, for which a small modification of the Stern-Gerlach force due to an admixture of the $m_F=0$ state from the Raman coupling was compensated for by a modification of the magnetic field gradient, instead of using an inertial force (Methods). **c**, Relative variation of the cloud width along the z-axis on the phase, as determined from the ratio $\left(\Delta z_{max}(t_{int}) - \Delta z_{min}(t_{int})\right)/\left(\Delta z_{max}(t_{int}) + \Delta z_{min}(t_{int})\right)$ of the sinusoidal fit, versus the momentum width of the atomic cloud. Here $\Delta z_{max}$ ($\Delta z_{min}$) denote the maximum (minimum) rms cloud

width at the used $t_{int}$=1.7ms lattice interaction time. The shown error bars correspond to the standard deviation of the cloud width variation and of the mean momentum width respectively. The solid green line is a simulation. The inset shows the calculated loading efficiency into the topological edge state (for $\varphi=\pi/2$) versus the momentum width.

**Methods:**

*Experimental apparatus and procedure.* A schematic of the used experimental setup is shown in Fig.6. The experiment takes place within an ultrahigh vacuum chamber, which for the suppression of ac magnetic field noise is placed within a single layer μ-metal shielding with optical access for the cooling, trapping, detection and optical lattice beams. The magnetic field gradient required for the generation of the amplitude-chirped optical lattice and for the compensation of the Earth's gravitational acceleration is produced by two pairs of identical coils in an anti-Helmholtz configuration oriented at an angle of 45° with respect to the vertical axis. This configuration in good approximation provides a two-dimensional quadrupole magnetic field of the form $\vec{B}(x,y,z) = b(z\vec{e}_x + x\vec{e}_z)$ (see Fig.6). The atomic cloud is placed ≈300μm below the center of the quadrupole field. At this position (with $x\approx0$ and $z>0$), the field gradient for rubidium atoms ($^{87}$Rb) in the F=1, $m_F$=-1 component of the electronic ground state causes a vertically upwards directed Stern-Gerlach force. For a gradient $b = mg/(\mu_B|g_F|) \cong 30.5 \text{G cm}^{-1}$, this compensates for the gravitational acceleration of rubidium atoms. Here $g_F = -1/2$ denotes the gyromagnetic factor for the used F=1 hyperfine component of the electronic ground state. Experimentally, the gravitational force can be compensated to within 1 part in $10^4$.

The used experimental setup is a modified version of an apparatus used in earlier works[19,24]. The experiment proceeds by loading cold rubidium atoms collected in a magneto-optical trap into the dipole trapping potential provided by a focused beam derived from a $CO_2$-laser operating near 10.6 μm wavelength. The atoms are evaporatively cooled by lowering the power of the mid-infrared trapping beam, which produces a Bose-Einstein condensate of ~15000 atoms in the F=1, $m_F$=-1 component of the electronic ground state. The magnetic field gradient from the quadrupole field here allows for the production of a spin-polarized atomic cloud, and the quadrupole field provides additional confinement of the atomic cloud along the optical dipole beam axis. To enhance the overlap with the topological edge state, the condensate cloud is directly after production adiabatically expanded by further lowering the mid-infrared dipole trapping beam within a 10s long ramp. During this expansion the effective trap size of the dipole trapping beam is effectively enlarged along the z-axis by acousto-optic rapid modulation of its focal position (with ≈50kHz modulation frequency). We arrive at an ensemble of ≈10000 atoms of $\Delta z \cong 11.5 \mu m$ rms spatial width along the lattice beam axis. The trapping frequency of the dipole potential along this axis at the end of this sequence is $\omega_z/2\pi \cong 4$ Hz. The rms momentum width of the trapped atomic cloud was determined to be $\Delta p_z \approx 0.01 \hbar k$ along the lattice beam axis, using Bragg spectroscopy performed immediately after extinguishing the dipole trapping potential. This value corresponds to an effective kinetic temperature of 35pK. Regarding the case of free atomic clouds, two-dimensional matter wave lensing experiments with a rubidium Bose-Einstein condensate have achieved effective temperatures of 50pK[28]. For the data points with higher momentum spread shown in Fig.5c, the adiabatic expansion of the condensate cloud proceeded to less low values of the trap potential, for which correspondingly also $\omega_z$ at the end of the ramp was higher.

The optical radiation required to generate both the optical Bragg pulses and the optical lattice is derived from a single mode high power diode laser operating near 783.5 nm wavelength, 3.3 nm detuned to the red of the rubidium D2-line. Its emission is split into two, and each of the beams pass an acoustic-optical modulator used to imprint different optical frequency components. They are then guided through optical fibers to the vacuum chamber, where the beams are irradiated in a vertically oriented, counterpropagating geometry onto the cold atomic cloud.

Following the adiabatic expansion of the atomic cloud, the dipole trapping beam is extinguished and the atoms are irradiated with two simultaneously performed Bragg pulses of opposite direction of the momentum transfer, tuned to transfer atoms in a state $\phi_a(z)$ to the superposition $\phi_i(z) = \frac{1}{\sqrt{2}} \phi_a(z)\left(e^{i2kz} + e^{-i\varphi}e^{-i2kz}\right)$, where φ denotes a variable relative phase. For more details on the Bragg-pulse technique, see Ref. 24.

To synthesize the amplitude-chirped lattice we use lattice potentials realized in a Raman-configuration (see Fig.2a of the main text), because the local depth (and amplitude) of such a lattice can be selectively controlled with magnetic fields of moderate magnitude. The local field value tunes the local value of the Raman detuning via the Zeeman shift. Four-photon processes, driven by a beam of frequency ω and two counterpropagating fields of frequencies ω+Δω and ω−Δω, respectively, (see Fig.6) here induce a lattice potential with a λ/4 spatial periodicity[18]. The spatial periodicity of the four-photon lattice is a factor 2 smaller than the λ/2 periodicity of a usual standing wave lattice induced by two-photon processes. Doppler-sensitive four-photon Raman transitions couple the momentum states $e^{i(2\hbar k+q)z/\hbar}$ and $e^{i(-2\hbar k+q)z/\hbar}$, respectively. Experimentally, the weak binding limit of the lattice potential is well fulfilled. For the two superimposed four-photon lattice potentials i=1,2 with opposite spatial variation of the lattice depth we choose values for the frequency shift of $\Delta\omega_1=\omega_z+\delta_0$ and $\Delta\omega_2=\omega_z-\delta_0$, respectively, where $\omega_z/2\pi \cong 700$kHz denotes the size of the Zeeman splitting between adjacent spin projections at z=0 and $\delta_0/2\pi \cong 200$kHz is the modulus of the Raman detuning at this spatial position. Both potentials are for atoms in the used $m_F$=-1 spin projection of the F=1 hyperfine ground state component. The two potentials $V_1(z)$ and $V_2(z)$ are generated using positive and negative two-photon detuning respectively and have an opposite spatial chirp of their corresponding lattice depth (as was shown in Fig. 2b of the main text). The potentials both have a spatial periodicity of $\lambda/4$, and by suitable choice of the phase values of the driving optical beams the potential maxima of say $V_1(z)$ are tuned to the position of the minima of $V_2(z)$, see Fig. 2b of the main text. The total potential experienced by the atoms $V(z) = V_1(z) + V_2(z)$ can be written in the form $V(z) = K(z)\cdot 2\cos(4kz)$. With the quoted experimental parameters, the coupling $K(z)$ is approximately linear in z in the experimentally relevant range (with the next term in the corresponding Taylor expansion ($\propto z^3$) reaching about 25% of the linear term at positions $z=\pm 50\mu m$).

Both during the evaporative cooling and during the time that the lattice is switched on, the magnitude of the field gradient is set by the requirement that the Stern-Gerlach force compensates the Earth's gravitational acceleration. However, besides imprinting the periodic lattice potential, the Raman coupling induced by the lattice beams cause a small admixture (≈2%) of the $m_F$=0 to the $m_F$=-1 Zeeman state. This reduces the effective magnetic moment, and thus the Stern-Gerlach force by a corresponding amount upon activation of the lattice beams. For the data shown in Figs.4, 5a, and 5c, and the data points shown by the blue dots in Fig.5b this was compensated for by a small chirp of the lattice beams eigenfrequencies. This induces a counteracting inertial force in the atomic frame. For comparison, the green squares shown in Fig.5b give the results for a second data set for which the applied magnetic field gradient was increased within 100μs upon activation of the lattice beams to compensate for the reduced magnetic moment of atoms. Here, no chirp of the lattice beams frequencies was required. Within experimental uncertainties the results of both data sets agree. The latter data set exhibited enhanced statistical uncertainties in comparison with the first one, as understood from the switching process. The experimental lifetime of the atomic cloud in the lattice potential is mainly limited by photon scattering from the optical Raman radiation. A further contribution from interatomic scattering channels has been estimated to be negligible, see below.

Detection of the atomic cloud following the interaction with the lattice was performed after extinguishing the lattice beams by activating both repumping light (tuned to the F=1 →F=2 component of the rubidium D2-line) and pulsing on a laser beam tuned to the F=2→F'=3 cycling component of this transition. This records a shadow image on a sCMOS camera (model Andor Zyla 5.5) in order to monitor the in site spatial distribution $n(y,z)$ of the atomic cloud. For the used magnification of the imaging system, the size of one pixel of the camera corresponds to 0.985 µm in the object plane. The size of the observed atomic cloud images along the $x$-axis (transversely to the axis of the lattice beams) is dominated by the spatial resolution of our imaging system, which in an independent measurement was determined to 4.8µm rms spatial width.

*Analysis*

To obtain the characteristic oscillation frequency $\omega_0$ from our experimental data, we have determined the (over $x$ and $y$) integrated atomic density $n(z) = \int_y n(y,z)$, corresponding to one-dimensional profiles, from the measured absorption images. Fig.7a shows such profiles $n(z)$ for atoms stored in the amplitude-chirped lattice for an interaction time $t$=1.7ms both for $\varphi = +\pi/2$ (top) and $\varphi = -\pi/2$ (bottom), corresponding to the data shown in Fig.4 in the main text for the corresponding interaction time. For the case of a phase $\varphi = -\pi/2$, we fitted a curve of the form $A_1 \exp(-(z - \delta z/2)^2 /(2\Delta z^2)) + A_2 \exp(-(z + \delta z/2)^2 /(2\Delta z^2))$ to the corresponding profile, which allows us to determine a value for the spatial splitting $\delta z$ between the two separating atomic clouds. Fig.7b shows the time evolution of the splitting $\delta z$ for different holding times in the optical lattice, which clearly indicates the onset of an oscillatory motion. The value of the characteristic frequency $\omega_0/2\pi \approx 158(16)$ Hz stated in the main text gives the obtained oscillation frequency of the sinusoidal fit. For the expected oscillation, see also Fig.8c,d and the corresponding discussion below.

*Theoretical background*

Assume that a rapidly oscillating potential $V(z) = 2K(z)\cos(4kz)$ with a smooth envelop function $K(z)$ (with $K(z) = az$ in our experiment) scatters particles from momentum $2k\hbar$ to $-2k\hbar$. Scattering to other momentum states is neglected in our treatment due to these states being energetically relatively far from resonance, see also Ref. 19. Parametrizing the wave function by $\phi(z,t) = (\Psi_1(z,t)e^{i2kz} + \Psi_2(z,t)e^{-i2kz})e^{-it\hbar(2k)^2/2m}$, the Hamiltonian $H_0 = \frac{p_z^2}{2m} + V(z)$ can be mapped to a one-dimensional Dirac equation for the spinor $\psi = (\Psi_1, \Psi_2)$ with

$$\hat{H}_D = -i\hbar c_{eff}\sigma_z\partial_z + K(z)\sigma_x \qquad (3)$$

where $\sigma_i$ are Pauli matrices and we have linearized the dispersion around $\pm 2k\hbar$ to obtain the effective speed of light $c_{eff} = 2\hbar k/m$. The Dirac equation possesses an emergent chiral symmetry

$$\sigma_y \hat{H}_D \sigma_y = -\hat{H}_D. \qquad (4)$$

According to Altland-Zirnbauer classification scheme[29], systems with this symmetry belong to the class BDI. In one dimension such systems possess non-trivial topological states which can be classified with an integer winding number[3]. Most importantly, states with $K(z) > 0$ and $K(z) < 0$ belong to two different topological classes where the difference of winding numbers is 1. Topology enforces the existence of a bound state with energy 0 given by

$$\psi(z) \propto e^{-\int_0^z K(z')dz'} \begin{pmatrix} 1 \\ -i \end{pmatrix} \quad \text{for } K(z \to \pm\infty) \gtrless 0$$ [30]. All other states come in pairs with energy $\pm E$ due to the chiral symmetry (4). For $K(z) = az$ the problem simplifies (and maps to a problem, exactly equivalent to the solution of the 2D Dirac equation in a magnetic field well-known from graphene[31]). The observation that $\hat{H}_D^2$ maps (up to a spin-dependent shift) to a harmonic oscillator, motivates the introduction of bosonic raising and lowering operators, $b^+, b$, as used for a standard harmonic oscillator. Thus, the Dirac Hamiltonian (3) is written in the following form

$$\hat{H}_D = \hbar\omega_0 \begin{pmatrix} 0 & b^+ \\ b & 0 \end{pmatrix} \tag{5}$$

where the matrix is given in the basis $|\uparrow\rangle = (1,-i)/\sqrt{2}$, $|\downarrow\rangle = (-i,1)/\sqrt{2}$ and $\omega_0 = \sqrt{2c_{\text{eff}}a/\hbar}$, and $\hat{H}_D^2 = (\hbar\omega_0)^2 \left(b^+b + \frac{1}{2} - \frac{1}{2}\sigma_z\right)$. The eigenfunctions of $\hat{H}_D$ are analytically described by

$$\psi_n^\pm = 1/\sqrt{2}\left(\pm\phi_n^{\text{ho}}|\uparrow\rangle + \phi_{n-1}^{\text{ho}}|\downarrow\rangle\right) \text{ for } n \geq 1, \text{ and } \psi_0 = \phi_0^{\text{ho}}|\uparrow\rangle \text{ for } n = 0, \text{ where}$$

$$\phi_n^{\text{ho}}(z) = \frac{1}{\sqrt{2^n n!}} \left(\frac{a}{\pi\hbar c_{\text{eff}}}\right)^{1/4} \exp\left(-\frac{az^2}{2\hbar c_{\text{eff}}}\right) H_n(\sqrt{a/\hbar c_{\text{eff}}}\, z)$$ are the eigenstates of a harmonic oscillator with oscillator length $\sqrt{c_{\text{eff}}\hbar/a}$ and Hermite polynomials $H_n(z)$. The spatial variation of the probability density for $n \leq 3$ was shown in Fig.1b of the main text and the energy spectrum of $\hat{H}_D$ is given by $E_n = \pm\sqrt{n} \cdot \hbar\omega_0$ with a unique zero energy eigenstate.

In Fig.8a we show that for the experimental parameters the description in terms of the Dirac equation is valid with high precision for $n \lesssim 20$. In the figure we compare the eigenvalues of the Dirac equation $H_D$ with the full spectrum obtained by diagonalizing the original Hamiltonian $H_0$.

*Interaction effects*

Two types of interaction effects have to be considered for our setup. First, interactions affect the initial state, i.e., the shape of the wave function before the application of the Bragg pulse. Second, they modify the time evolution after the pulse. We will first show that the second effect is negligible while the first has to be taken into account.

To estimate the importance of interaction effects for the time-evolution quantitatively, we consider a simplified one-dimensional situation where we assume that in perpendicular direction only the lowest energy states of a harmonic oscillator with frequencies $\omega_x/2\pi = 28.5\text{Hz}$ and $\omega_y/2\pi = 10\text{Hz}$ are occupied. This approximation actually overestimates interaction effects as it underestimates the size of the wave function in perpendicular directions. After projection onto degrees of freedom described by the Dirac spinor $\psi = (\Psi_1, \Psi_2)$, one obtains a one-dimensional Dirac-equation with local interaction and a corresponding Gross-Pitaevskii-Dirac equation which takes the form

$$i\hbar\partial_t\psi = H_D\psi + g\hbar c_{\text{eff}}\left(3|\psi|^2 - (\psi^+\sigma_z\psi)\sigma_z\right)\psi \qquad (6)$$

with the dimensionless interaction strength given by $g \approx N\dfrac{a_s m\sqrt{\omega_x\omega_y}}{2\hbar k}$, where we use the convention $\int|\psi|^2 = 1$. $N \approx 10^4$ is the number of particles, $a_s \approx 100 a_0$ is the scattering length of $^{87}$Rb, where $a_0$ describes the Bohr radius, and $\omega_x, \omega_y$ is the trapping frequency in the x- and y-direction, respectively. We obtain for our setup $g \approx 0.5$. This factor parametrizes the ratio of the interaction energy and the kinetic energy. It is important to note that the dynamics of the condensate is therefore *not* dominated by interactions (in contrast to the initially prepared cloud, which had a kinetic energy which was two orders of magnitude smaller). The dominant effect of $g$ is a shift of all energies linear in $g$, but all observables studied in our experiment are affected only to quadratic order in $g$ as a consequence of the chiral symmetry

4of $H_D$. A quantitative analysis shows that all interaction effects turn out to be only on the percent level and thus negligible within our present experimental resolution.

To study the interaction effects, we have solved both the stationary and the time-dependent Gross-Pitaevskii-Dirac equation and, furthermore, calculated the Bogoliubov spectrum. Fig.8b shows that the shape of the stationary solution of the Gross-Pitaevskii-Dirac equation for *n*=0 is almost unaffected by the interaction. Similarly, the density oscillations of the superposition of the doublet with n=1 are indistinguishable for $g = 0.5$ and $g = 0$, see Figs.8c and d.

An important qualitative effect of interactions in the final state is that the system is intrinsically unstable. While for fermions Pauli blocking prohibits the decay of the topological surface state, the bosonic condensate wave function is not stable in the presence of interactions. As a consequence, the energies of the Bogoliubov spectrum describing the fluctuations around the stationary solution of the Gross-Pitaevskii-Dirac equation obtain imaginary contributions. An explicit calculation shows, however, that all decay rates are smaller than 1% of the typical oscillation frequency $\omega_0$.

While interaction effects thus do not play a role for the time evolution in our experiment, they do affect the initial state substantially. The main reason for this is that before the Bragg pulse is applied the kinetic energy is about 2 orders of magnitude smaller than after the pulse. Therefore, interaction effects dominate the initial, but not the final state. To gain a quantitative theoretical description of the experiment (lower panels in Fig.4, green curve in Fig.5c), we therefore proceed in the following way. To obtain the initial state wavefunction (after adiabatic expansion, before the Bragg pulse and external lattice is switched on) we first calculate the static solution $\psi_a(x, y, z)$ of the 3-dimensional Gross-Pitaevskii equation using

Ref. 32. After the Bragg pulse (and after switching on of the lattice potential), the wavefunction is described by $(\psi_a e^{2ikz} + \psi_a e^{-i\varphi} e^{-2ikz})/\sqrt{2}$. For the subsequent time evolution, we ignore interaction effects and use the non-interacting Dirac Hamiltonian (3) in combination with simple harmonic oscillator Hamiltonians for the perpendicular directions. The experiment determines the cloud from an absorption image in the *x-z*-plane. In the lower panel of Fig.4 we therefore plot $\int dy |\Psi_1(x,y,z,t)|^2 + |\Psi_2(x,y,z,t)|^2$ convolved with a Gaussian of rms width 4.8μm to describe the effects of the finite experimental resolution. Note that the experimental resolution has little effect for the dynamics in *z*-direction and mainly broadens the perpendicular direction. Rapidly oscillating interference fringes arising from $2\text{Re}[\Psi_1(x,y,z,t)^* \Psi_2(x,y,z,t) e^{-i4kz}]$ cannot be resolved within the experimental resolution.

To obtain the green curve in Fig.5c, we have repeated the above described calculations for several values of $\omega_z$, which gives us initial atomic distributions with different values of the momentum width. The quoted values for the momentum width are determined from $\Delta p_z = \sqrt{\langle p_z^2 \rangle}$ in the initial state, the corresponding width of the final oscillating state from the value for $\sqrt{\langle z^2 \rangle}$, as obtained after interaction with the lattice.

**Figure 6 | Overview of cold atoms setup.** An optical dipole potential for evaporative cooling of rubidium atoms is generated by a focused mid-infrared beam derived from a $CO_2$-laser. The shown field coils at the position of the atomic cloud (~300μm below the center of the generated magnetic quadrupole field) create a vertically oriented magnetic field gradient used to imprint an amplitude-chirp of the four-photon lattice. The corresponding Stern-Gerlach force is counter-directed to the Earth's gravitational acceleration. Shown are the optical frequency components in the vertically oriented lattice beams used to synthesize the two superimposed four-photon lattice potentials (i=1,2).

**Figure 7 | Eigenstates and Analysis.**

**a,** Integrated atomic density distribution $n(z)$ for atoms stored in the lattice for $t=1.7$ms with the initial phases $\varphi = \pm\pi/2$. In the case of $\varphi = \pi/2$ the distribution has almost a Gaussian shape, while for $\varphi = -\pi/2$ it has a double peak structure. Each data point correspond to the average of four realizations, and the error bars are the standard deviation of the mean. **b,** From a sinusoidal fit to the variation of the splitting $\delta z$ of the two peaks during the evolution of the atoms in the lattice initially prepared with phase $\varphi = -\pi/2$ the characteristic frequency is obtained. Each data point correspond to the average splitting of four realizations, and the error bars are the standard deviation of the mean.

**Figure 8 | Validity of Dirac Equation.**

**a**, The eigenvalues of the microscopic Hamiltonian $H_0$ (circles) close to the band crossing agree with a precision higher than 1% with the energies (crosses) of the Dirac equation. **b,** Density profile of the bound state $n=0$ with (dashed line, $g = 0.5$) and without (solid line, $g = 0$) interactions. Changes of the density profile (inset) are of the order of 1%, much smaller than the experimental resolution. **c,d,** Time dependence of the density profile calculated from the solution of a Gross-Pitaevskii Dirac equation with and without interactions in the case of an initial phase of $\varphi = -\pi/2$. Here predominantly the doublet with *n*=1 is populated, and the probability distribution shows a temporal oscillation. On the time scale of the experiment interactions have no observable effect.

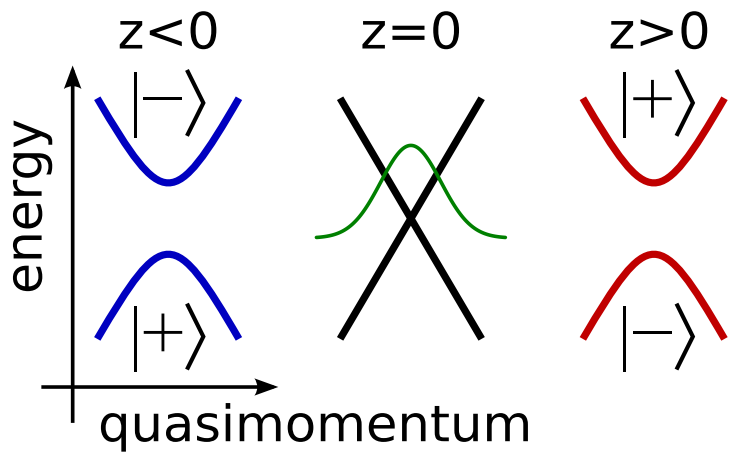

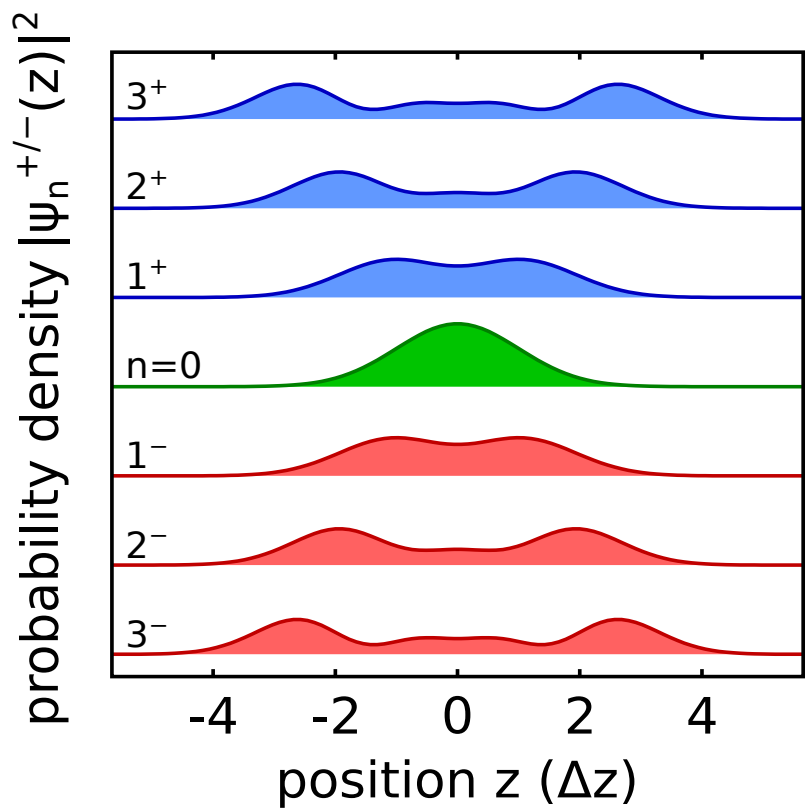

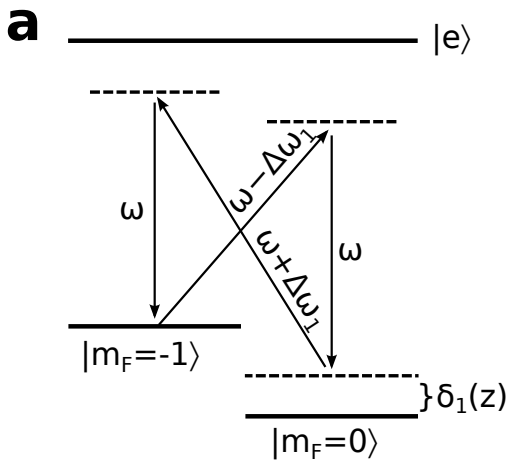
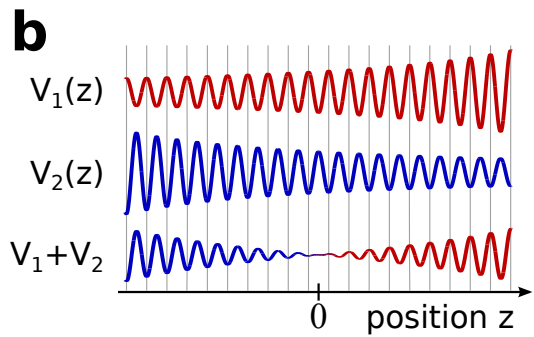
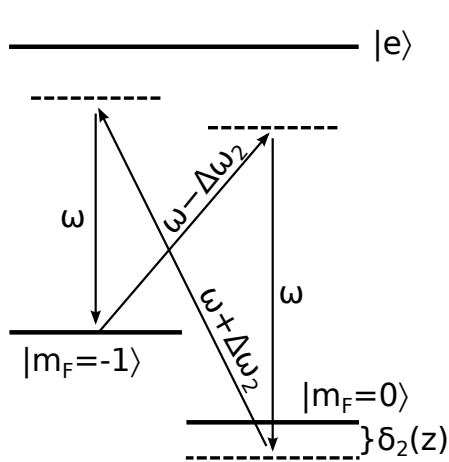
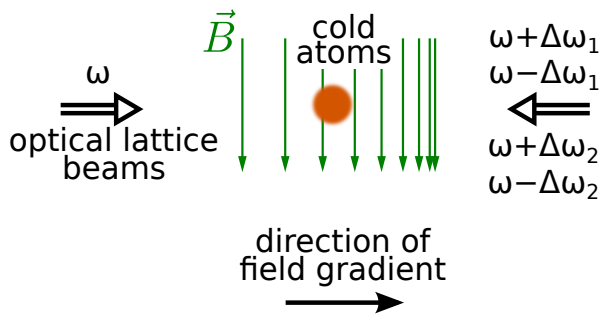

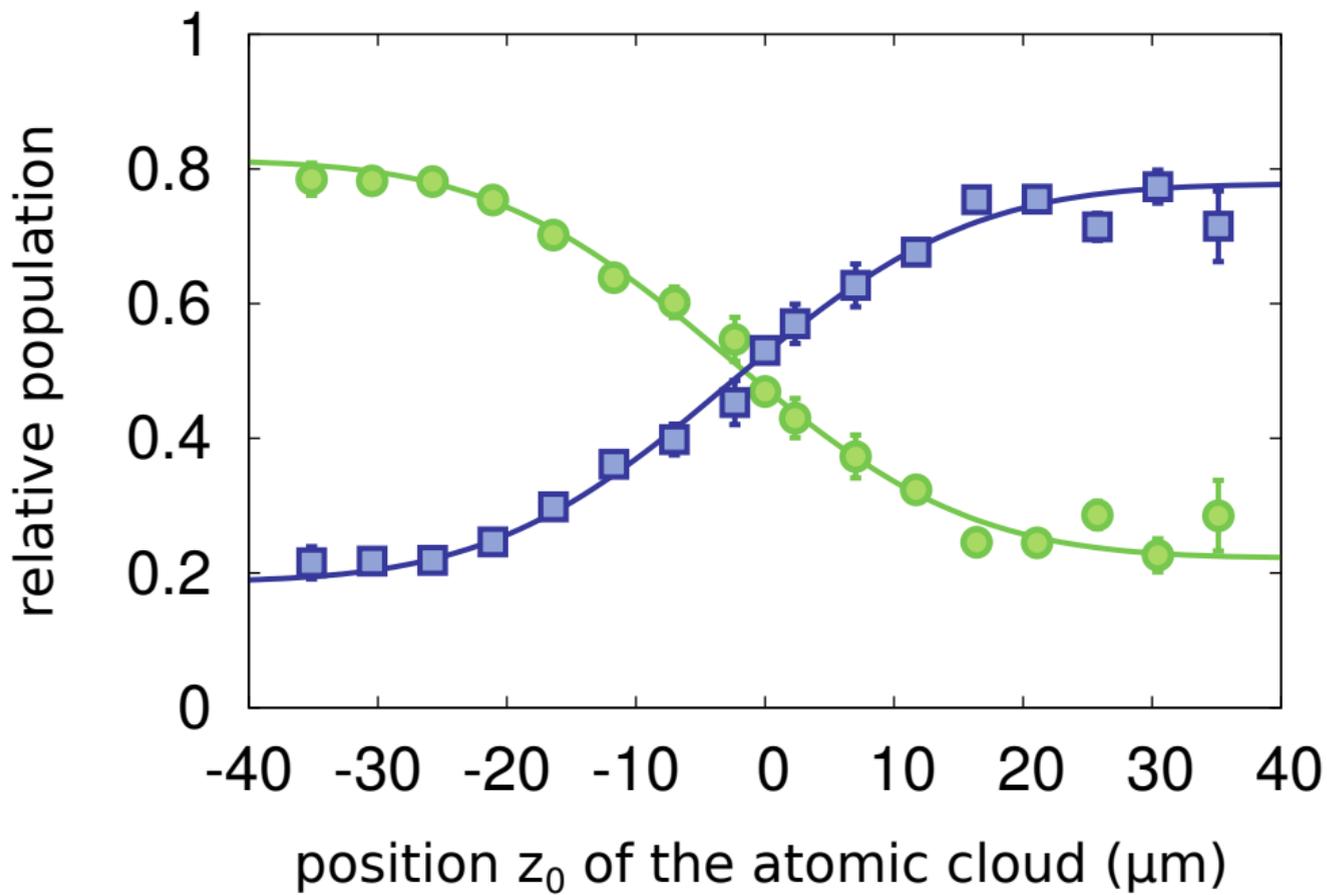

| **a** | **b** |
|---|---|
| φ = +π/2 | φ = −π/2 |

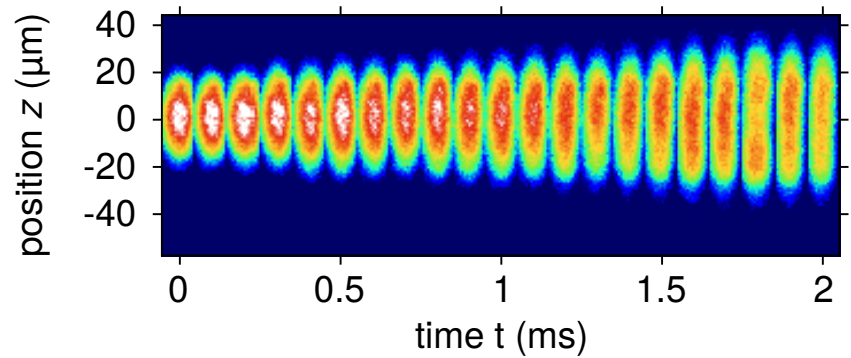 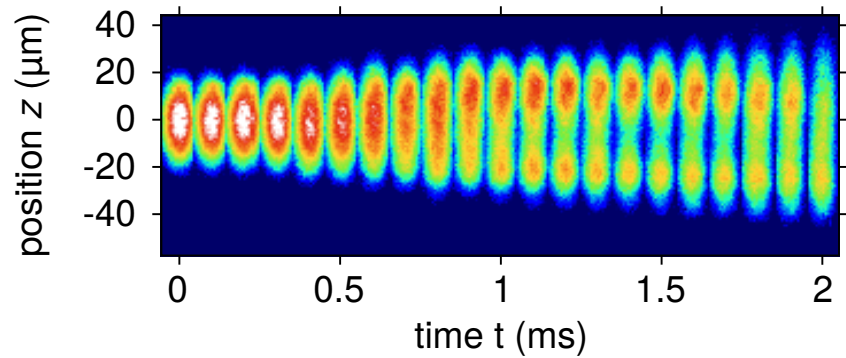

experimental data

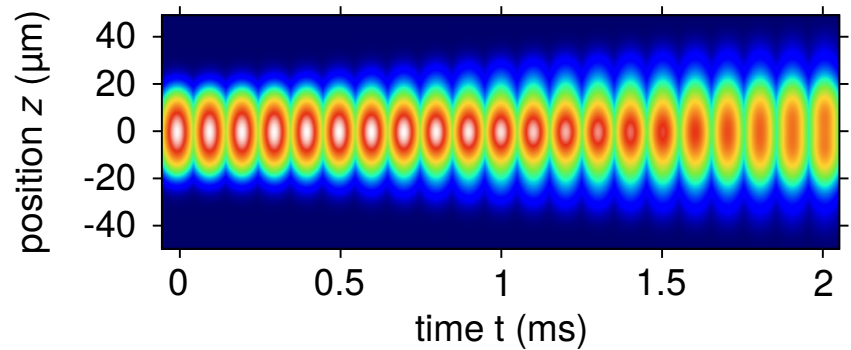 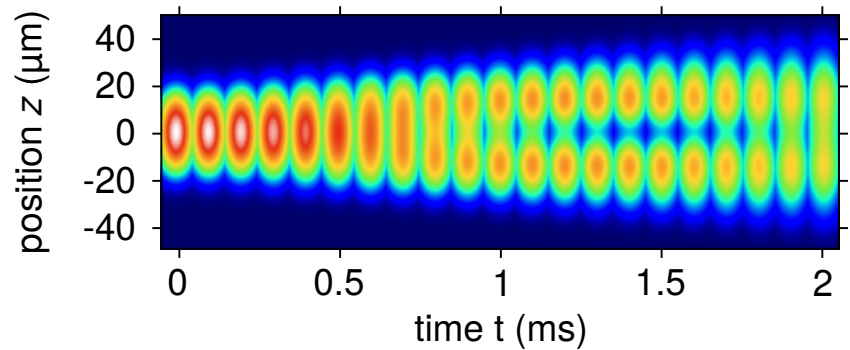

simulation

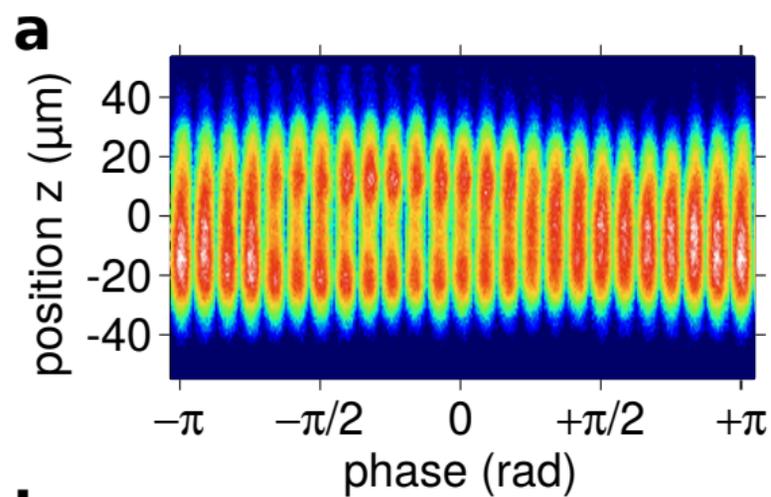
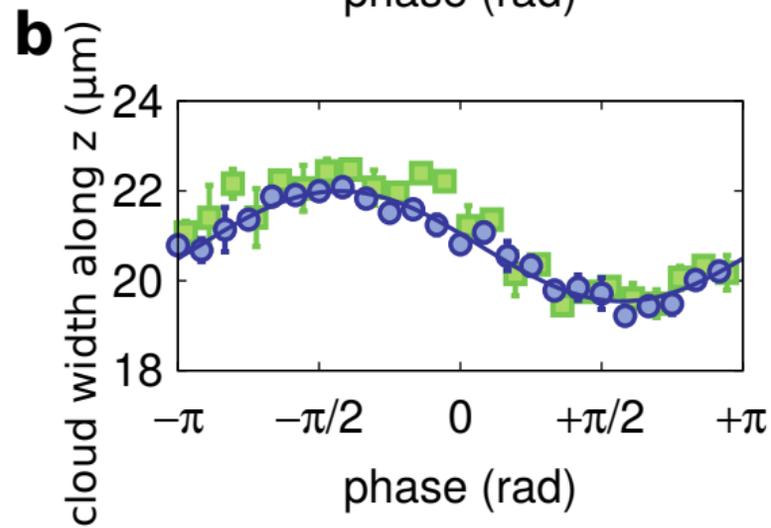
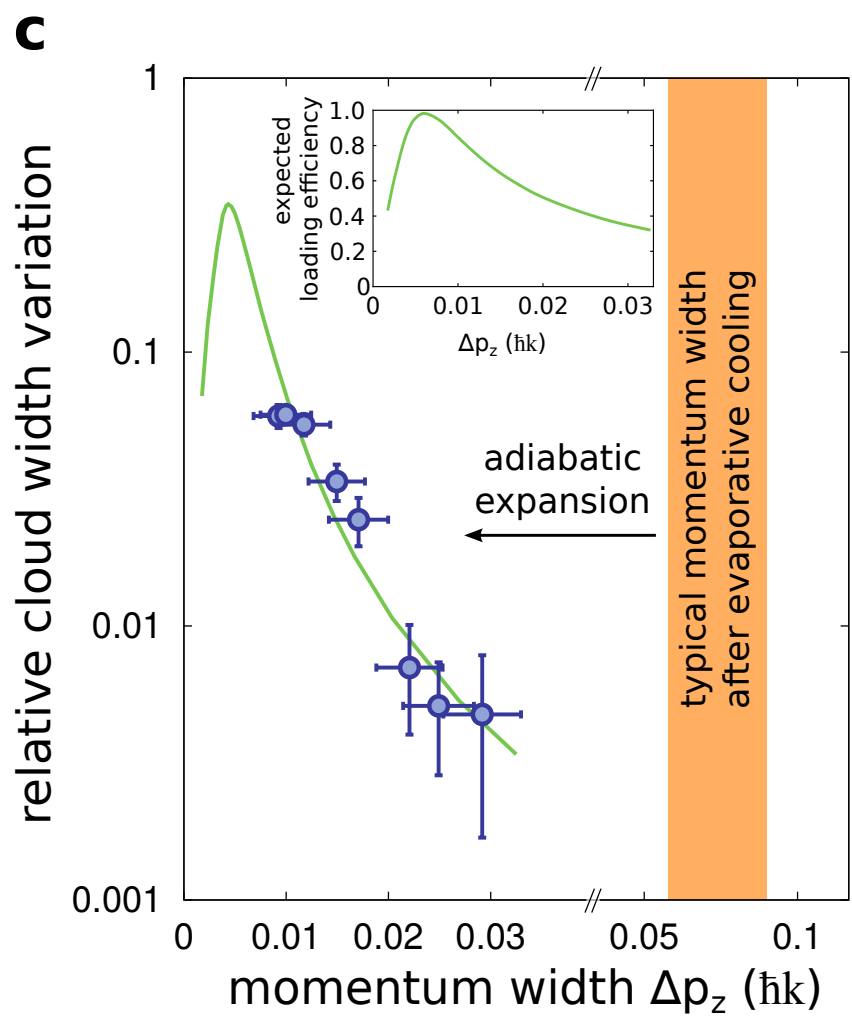

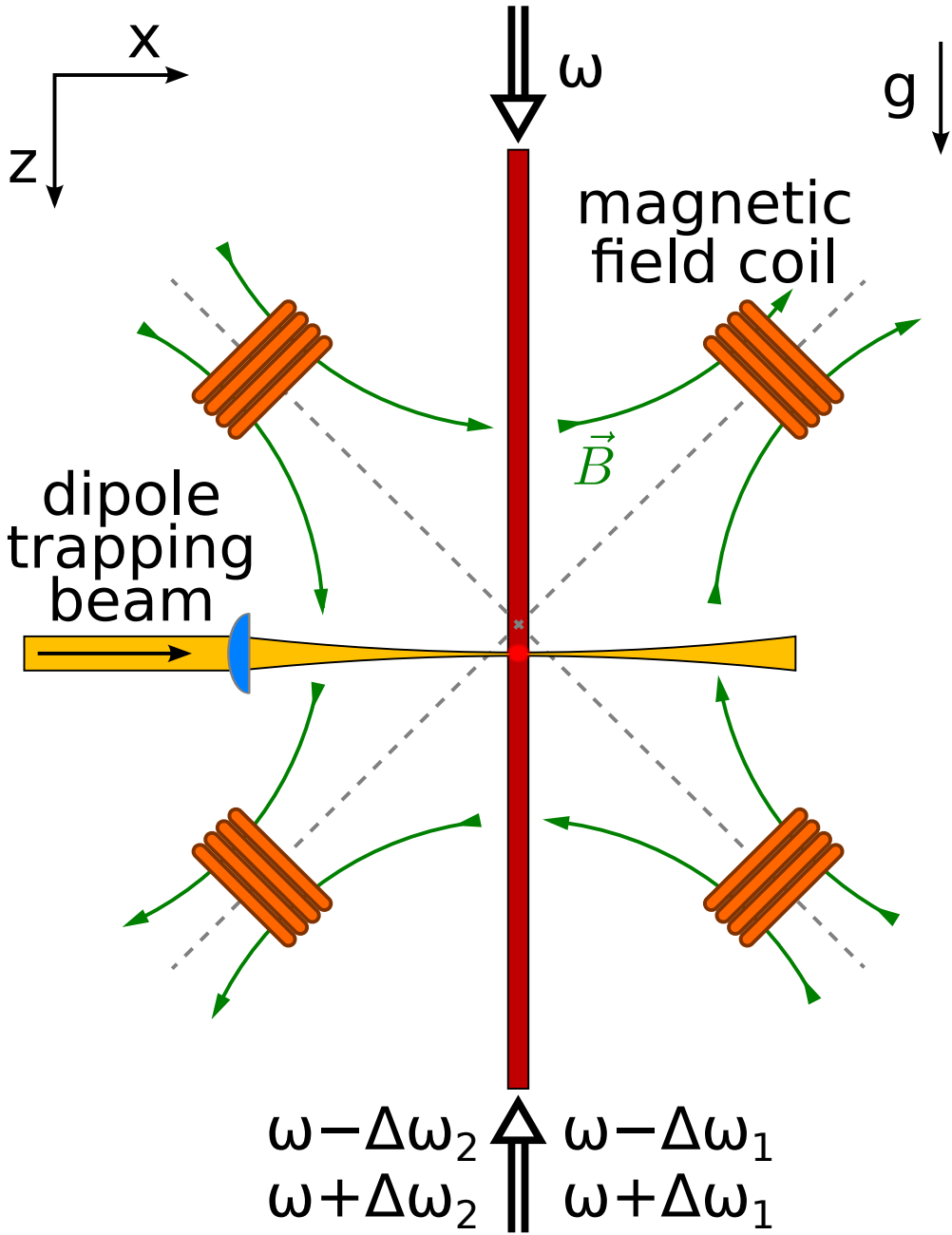

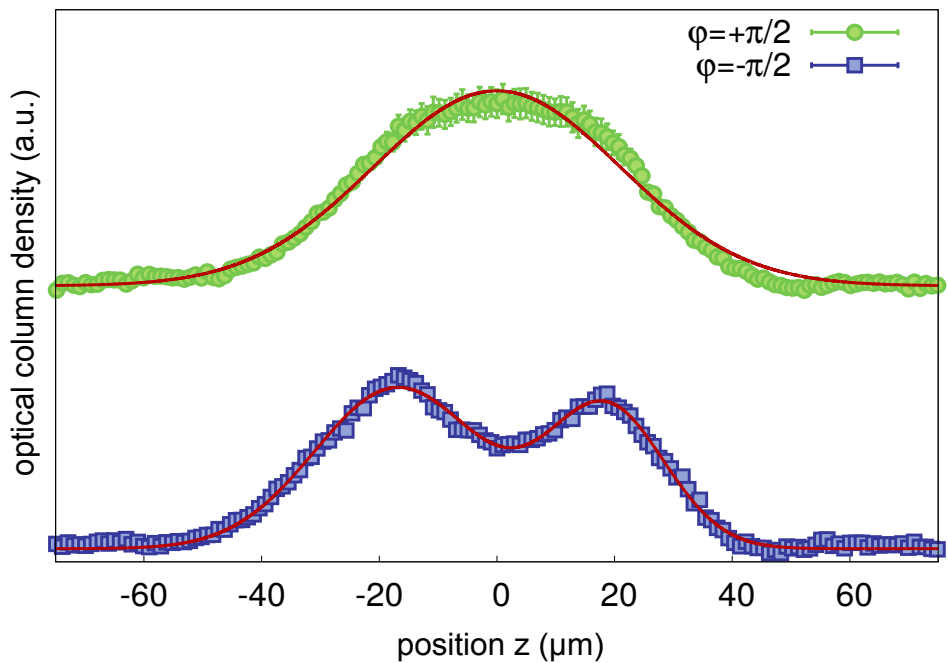

**a**

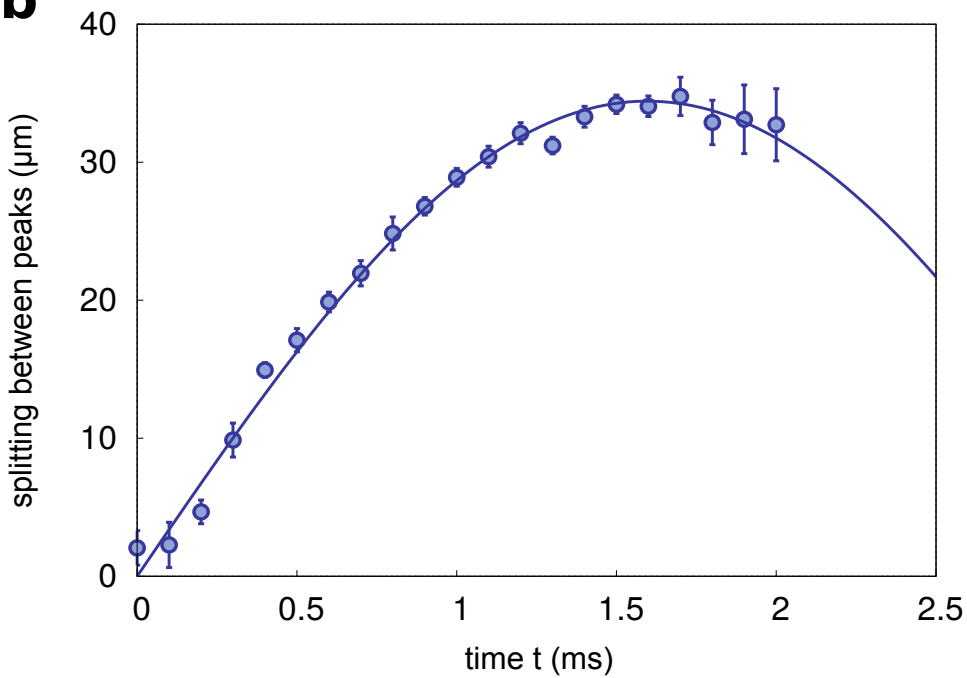

**b**

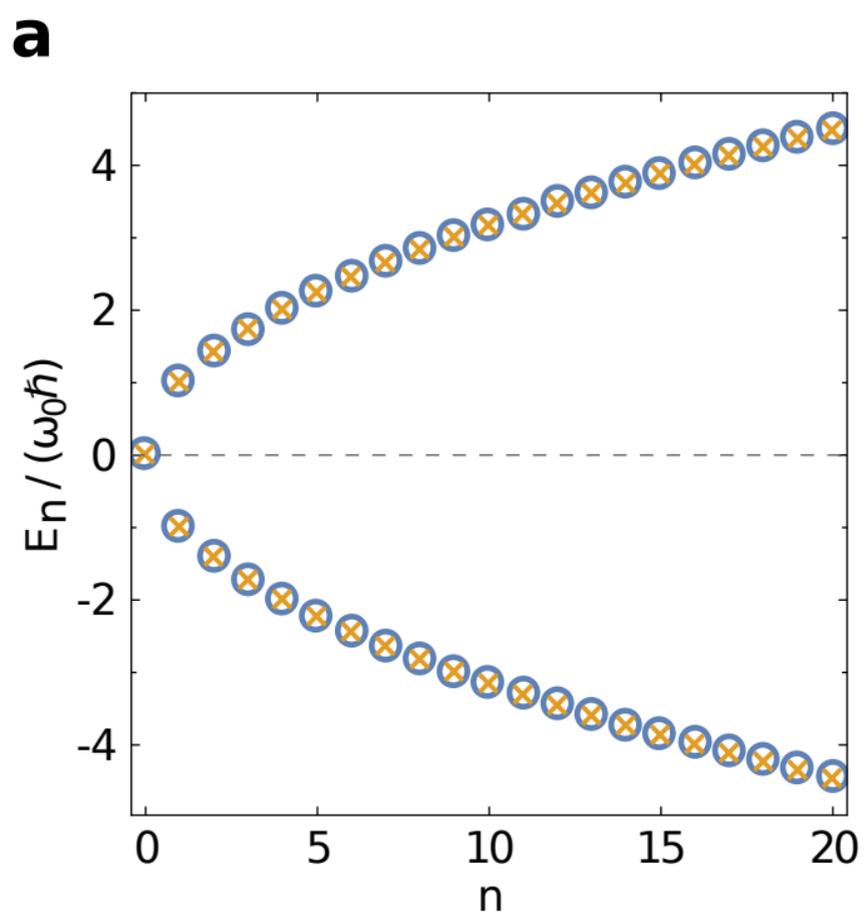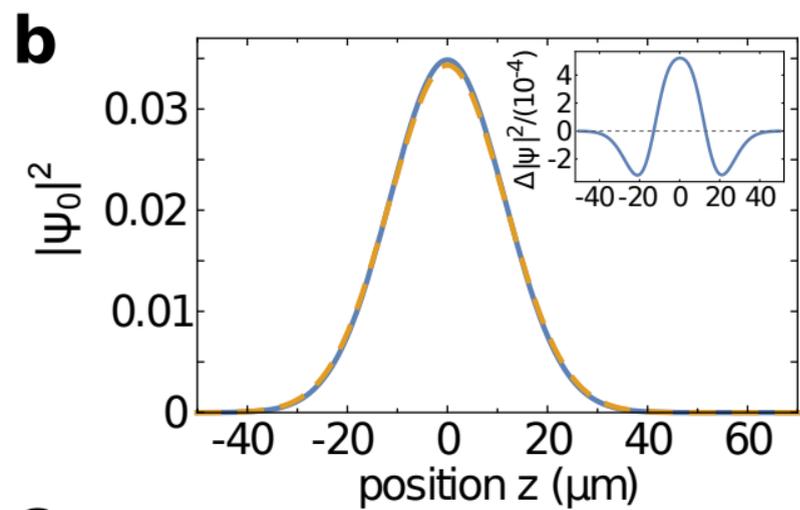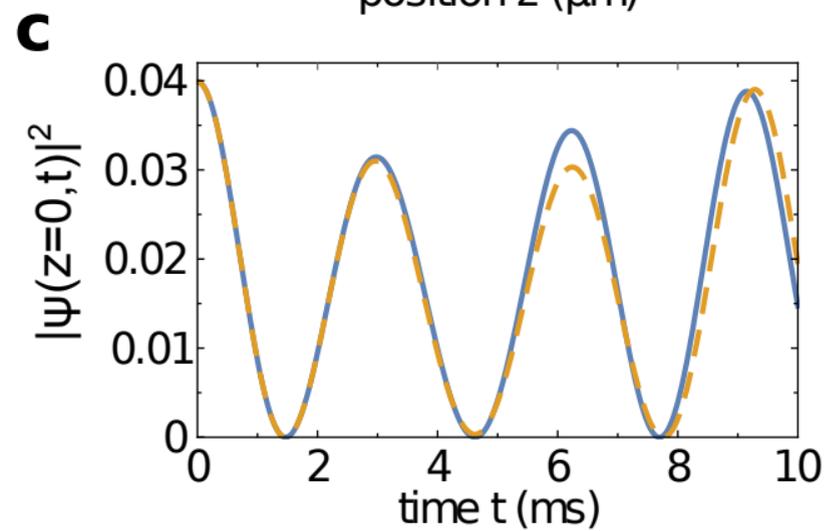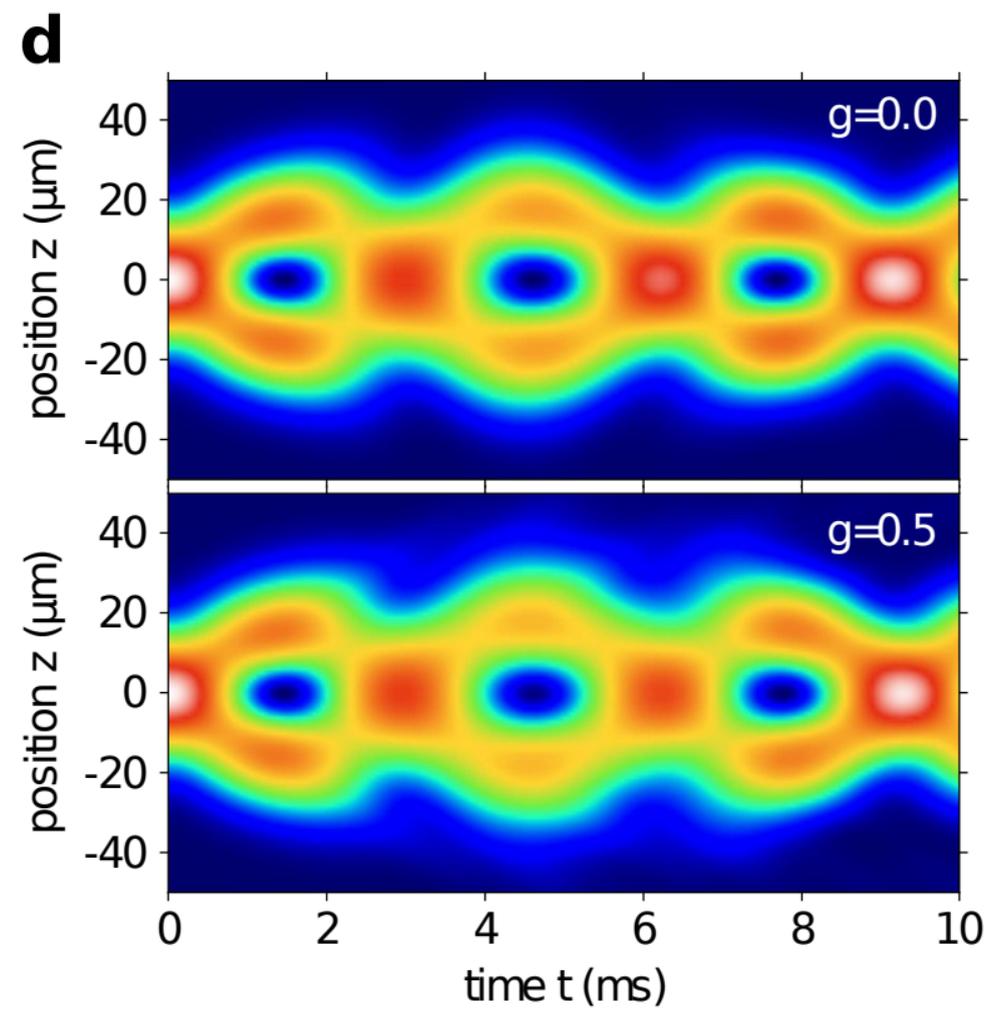